\documentstyle[sprocl,epsf]{article}

\def\ra{\rightarrow}

\def\be{\begin{equation}}
\def\ee{\end{equation}}
\def\bea{\begin{eqnarray}}
\def\eea{\end{eqnarray}}

\begin{document}
\begin{titlepage}

\vspace*{.3cm}

\begin{flushright}
TPI-MINN-31-97\\
HEP-97-1617\\
hep-ph/9712323
\end{flushright}

\vspace{1cm}

\begin{center}
{\bf HQET AND SEMILEPTONIC FORM FACTORS }
\\~
\\~\\
ARKADY VAINSHTEIN \\
{\em
Theoretical Physics Institute, University of Minnesota\\and\\
Budker Institute of Nuclear Physics, Novosibirsk\\~\\~\\
Talk at the Seventh International Symposium On Heavy Flavor Physics,\\ 
July 7-11,  1997, Santa Barbara;\\ to appear in the Proceedings}
\end{center}

\vspace*{2cm}

\begin{center}
{\bf Abstract}
\end{center}

\vspace*{.2cm}
\noindent
Theoretical approaches to form factors of semileptonic
decays  are discussed in application to $B \ra D + l+\bar \nu_l$ decay. 

\end{titlepage}

\title{HQET AND SEMILEPTONIC FORM FACTORS }

\author{ ARKADY VAINSHTEIN }

\address{Theoretical Physics Institute, University of Minnesota\\and\\
Budker Institute of Nuclear Physics, Novosibirsk}

\maketitle\abstracts{Theoretical approaches to form factors of semileptonic
decays  are discussed in application to $B \ra D + l+\bar \nu_l$ decay. }

\section{Outline}
Heavy Quark Effective Theory (HQET)~\cite{HQET} is a powerful approach to 
the processes involving heavy
flavor hadrons and, in particular, to the study of form factors of 
semileptonic decays (see 
reviews~\cite{Neub}).
From a field-theoretical standpoint, the HQET is a particular example of construction based
on a more general notion of Wilson  Operator Product Expansion (OPE) or Wilson effective
action~\cite{wilson}, to which I mainly refer.

The outline of the talk is as follows:
\begin{itemize}
\item{
 Heavy flavor decays as short distance probes. Total semileptonic widths}
\item{
 From inclusive to exclusive: bounds for $B\ra D^*$ leading form factor}
\item{
Subleading $B\ra D^*$  form factors}
\item{
 Some recent developments}
\item{
Conclusions}
\end{itemize}

\section{Heavy Flavor Decays as Short Distance Probes}

The quark picture of semileptonic $B$ meson decay $B \ra D + e+\bar \nu_e$ is presented by
Fig.~1. (Better to put  ${\bar B}$ but I will systematically neglect by bar on $B$.)
\begin{figure}[ht]
\epsfxsize=10.0cm
\centerline{\epsfbox{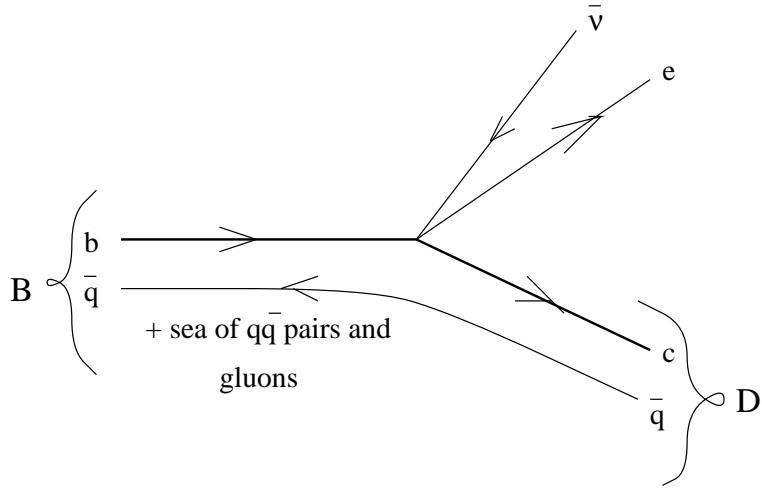}}
\caption{Semileptonic  $B \ra D + e+\bar \nu_e$ decay}
\end{figure}
The specific feature  of heavy flavor states is that they contain a heavy quark $Q$
which could be viewed as an almost static object (in the rest frame of hadron). Thus these
states can be considered as ground states of light flavor QCD in the presence of static source of
gluon filed. A relevant theoretical parameter is
\be
\frac{\Lambda_{QCD}}{m_Q} \ll 1\;.
\label{hm}
\ee
An immediate consequence is so called spin-flavor symmetry, i.e., an independence on  heavy
quark spin  and flavor  in the large mass limit where the relation (\ref{hm}) holds.

However, the characteristic size of heavy flavor hadron is of order $1/ \Lambda_{QCD}$  and
the structure of  heavy flavor states  as well as that of the QCD 
vacuum is governed by nonperturbative dynamics in strong coupling regime. 
 To see a simple underlying  quark-gluon structure at short distances, where the strong
interaction becomes  weaker one needs a short distance probe. Heavy  flavor decays, particularly
the ones inclusive over final hadrons, provide such a probe. This probe is similar to the famous
ratio
$R$ for the inclusive cross  section of hadron production with a substitution of QCD vacuum
by  a heavy flavor hadron. This comparison is illustrated by Figs.~2 and 3.
\begin{figure}[ht]
\epsfxsize=10cm
\centerline{\epsfbox{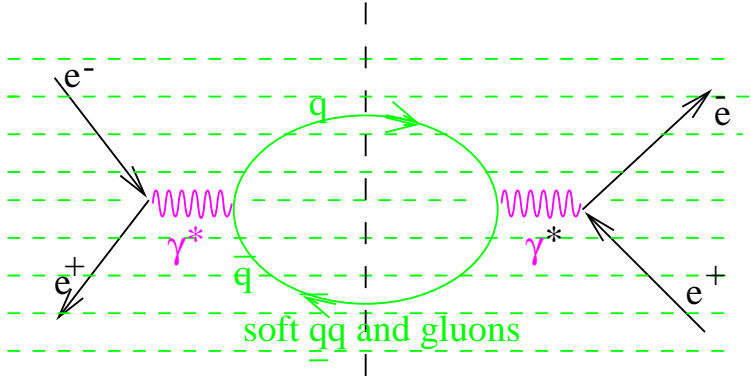}}
\caption{$e^+ e^-$ annihilation into hadrons (inclusive)}
\end{figure}
Fig.~2 is for the process of $e^+e^-$ inclusive annihilation into hadrons, its cross section is
given by 
\be
R=\frac{\sigma(e^+ e^- \rightarrow {\rm hadrons})}{\sigma(e^+ e^- \rightarrow
\mu^+
\mu^-)} = N_{{\rm colors}}\sum_q Q_q^2 \left[1+ O [\alpha_S (W)] + \dots \right],
\label{ee}
\ee
in the limit of high total energy $W$, i.e., when  $W\gg \Lambda_{QCD}$.
\begin{figure}[ht]
\epsfxsize=10cm
\centerline{\epsfbox{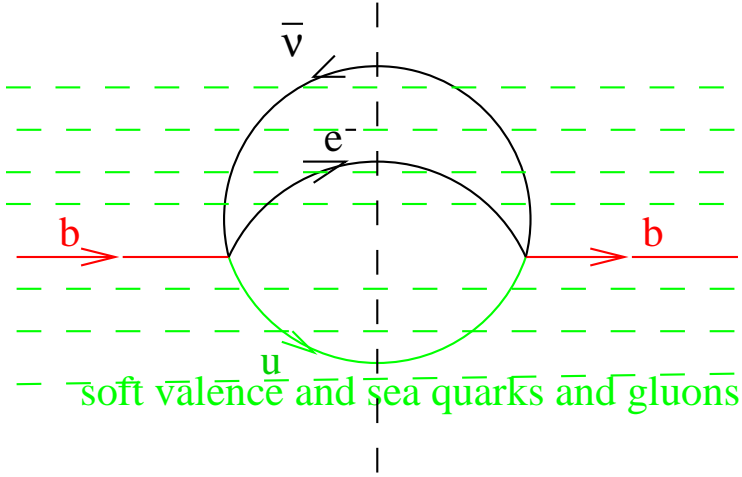}}
\caption{Inclusive width of $B \ra X_u e {\bar \nu}$ decays}
\end{figure}
Fig.~3 presents  semileptonic decays of $B$ meson inclusive over  nonstrange hadrons in the
final state ($b\ra u$ transition in terms of quarks), 
\be
\Gamma_{SL}^{ub}=\sum_{X_u}\Gamma (B \rightarrow X_u  e {\bar \nu})=
\frac{G_F^2}{192 \pi^3}
m_b^5 |V_{ub}|^2\,\left[
1+ O [\alpha_S (m_b)] + \dots \right]
\label{sl}
\ee 
in the limit of large mass of b quark, $m_b\gg \Lambda_{QCD}$.

In both examples the leading term is given by partonic description of the process 
where only hard   momenta count and an influence 
of soft background of quarks and gluons in the vacuum  and in B meson is strongly
suppressed.  Theoretical basis of quark-hadron duality is not trivial and based on two notions:

\vspace{0.2cm}

\noindent
(i) Wilson Operator Product Expansion~\cite{wilson}. The expansion takes place at large {\em
Euclidean}  momenta where no hadron can be produced. The OPE means  a factorization of
short and large distances in Euclidean domain.

\vspace{0.2cm}

\noindent
(ii) Dispersion relations. Based on causality  dispersion relations express certain integrals
over observable cross sections (in {\em Minkowski} domain) as correlators at  Euclidean
momenta.

\vspace{0.2cm}

The construction described allows for calculation of corrections to the leading partonic term, both 
perturbative, $[\alpha_S (m_b)]^n$, and nonperturbative, $[\Lambda_{QCD}/m_b]^k$.
For the semileptonic total width (\ref{sl}) it results in
\begin{equation}
\Gamma_{SL}^{ub}=\frac{G_F^2 m_b^5 |V_{ub}|^2}{192 \pi^3}
 \left[1- \frac{2\,\alpha_S}{3\pi}-\frac{\mu^2_\pi}{2m_b^2} -
\frac{3\,\mu^2_G}{2 m_b^2}
\right]
\label{btou}
\end{equation}
where power corrections~\cite{buv} are defined by parameters $\mu^2_\pi$ and $\mu^2_G$.
The first one, $\mu^2_\pi$, has the meaning of mean value for momentum square of $b$
quark inside of $B$ meson,
\be
\mu^2_\pi = \langle B|  {\bar b} {\vec \pi}^2 b |B\rangle \approx 0.6 \pm 0.1 \;{\rm GeV}^2\;.
\ee
The value given is due to theoretical estimate~\cite{bb}, the direct determination  is  not done
yet. The second parameter, $\mu^2_G$, measures chromomagnetic interaction with the spin
of heavy quark, so its value  is fixed by spin splitting in $B,\;B^*$ system,
\begin{equation}
\mu_G^2=\langle B|{\bar b}\frac{i}{2} \sigma^{\mu\nu} G_{\mu\nu} b
|B\rangle =\langle {\vec \sigma}{\vec B} \rangle
=\frac{3}{4}(M_{B^*}^2 - M_B^2) \approx 0.36\,{\rm GeV}^2\;.
\label{mugnum}
\end{equation}

\section{From  Inclusive to Exclusive}

The decays $B\ra D^* l {\bar \nu_l}$ and $B\ra D l {\bar \nu_l}$ produce about 2/3 of
the total semileptonic width. The suppression of production of higher states can be understood
as due to a closeness to the point of zero recoil. In other words, hadrons in the final state are slow
moving. The point of zero recoil, where the spatial momentum ${\vec q}=0$, is a special  one
in the case of heavy quark in the final state.  The quark-hadron
duality takes place at this point in spite of  small momentum transfer.
 It is a remarkable  case where  OPE works up to large distances (in case of light
quarks a similar phenomena shows up in the axial current anomaly and in the certain term
of OPE for the product of two axial currents).

The theoretical knowledge of the amplitude of $B\ra D^* l {\bar \nu_l}$ decay at zero
recoil is used, in particular, for an extraction of $|V_{cb}|$ from the data. The crucial
 for this extraction question is about a theoretical accuracy of HQET predictions.

Let us start with kinematics of the decay. There are four invariant form factors describing
the transition~\cite{Neub},
\bea
{\langle D^*|{\vec V} - {\vec A}|B\rangle}&= &\sqrt{m_B m_{D^*}}\, h_{A_1}(w) 
\left[ -(w+1) {\vec e}_{D^*}+ \right. \nonumber\\
& & \left. wR_1(w) i {\vec v}_{D^*} \times e_{D^*} +wR_3(w) {\vec v}_{D^*}({\vec v}_{D^*}
\cdot{\vec e}_{D^*})\right]\;,
\label{vecV}
\eea
\bea
{\langle D^*| V_0 -  A_0|B\rangle}&= &\sqrt{m_B m_{D^*}}\, h_{A_1}(w)\,
 ({\vec v}_{D^*}\cdot{\vec e}_{D^*})
\left[-(w+1) + ~~~~~~~~\right. \nonumber\\
& & \left. R_2(w) \,\frac{m_B}{m_{D^*}} -R_3(w) \left(\frac{m_B}{m_{D^*}}-1\right)\right]\;,
\eea
where matrix elements of $V-A$ current are written separately for spatial  
${\vec V}-{\vec A}$ and timelike  $V_0-A_0$ components and 
\be
w=\frac{E_{D^*}}{m_{D^*}}=\sqrt{1+\frac{{\vec q}^2}{m_{D^*}^2}}\;,~~~~~~~
{\vec v}_{D^*}=-\frac{{\vec q}}{m_{D^*}\,w}\;.
\ee
The $ h_{A_1}(w)$ form factor is the only one contributing at zero recoil. Subleading 
(in terms of their contribution to the amplitude  near zero recoil) form factors are
introduced in the form of their ratios
$R_i$ to the leading one. In the HQET limit  these ratios go 1,
\be
R_1(w)=R_2(w)=R_3(w)=1 ~~~~~~ ({\rm at }~~m_b,m_c\ra \infty)
\ee

\subsection{Zero Recoil Sum Rule}
The amplitude at zero recoil is defined by $ h_{A_1}(1)$ as discussed above.  To get its value
let us write down the sum rule~\cite{suv} for {\em inclusive} 
transitions $B\ra X_c l {\bar \nu_l}$ at zero recoil, ${\vec q}=0$,
\be
|h_{A_1}(1)|^2 + \sum_{m_{D^*}<m_{X_c}<m_{D^*}+\mu} |h_{n}(1)|^2 \,=\, \xi_A(\mu)
-\Delta_{1/m^2} + O\left( \frac{1}{m^3}\right)\;.
\label{sr}
\ee
The states $X_c$ produced at zero recoil are $D^*$ and its radial
excitations and
 $h_{n}(1)$ are analogs of $h_{A_1}(1)$ for radial excitations. The production of
excitation is suppressed in the heavy quark limit, $h_{n}(1)\sim 1/m$ and the sum in the l.h.s.
 of Eq.~(\ref{sr}) is of
$1/m^2$ order. At the r.h.s. the quantity
$\xi_A(\mu)$ gives a perturbative part and has the form
\be
\xi_A(\mu)=1 +2\frac{\alpha_S}{\pi}\left[ \frac{m_b+m_c}{m_b-m_c}
\ln\frac{m_b}{m_c}- \frac{8}{3}\right]+
\frac{2}{3}\frac{\alpha_S}{\pi} \mu^2\left[ \frac{1}{m_c^2}+\frac{1}{m_b^2}
+ \frac{2}{3m_b m_c}\right]
\ee
Let me note here that the full calculation of the $\alpha_S^2$ corrections was
 recently performed~\cite{cmu}. It
turned out to be a small change as compared with  the value taken from the BLM procedure.
Numerically the result for $\xi_A$ is
\be
 \xi_A(\mu=0.5 m_c)=0.99 \pm 0.02
\ee

The quantity $\Delta_{1/m^2}$ presents nonperturbative effects which are of the
second order in $1/m$, Eq.~(\ref{sr}) contains no corrections of the first order in $1/m$ in
correspondence with Luke theorem~\cite{Luke},
\be
\Delta_{1/m^2}=\frac{\mu_G^2}{3m_c^2}+ \frac{\mu_\pi^2-\mu_G^2}{4}\left[
\frac{1}{m_c^2}+\frac{1}{m_b^2} + \frac{2}{3m_b m_c}\right]\;.
\ee
Numerically it is about 0.1 which is not a small effect, particular for the quantity of the
second order in $1/m$. 

The most uncertain in the use of sum rule (\ref{sr}) for theoretical evaluation  of
$h_{A_1}(1)$ is the sum over excitation. In the absence of direct experimental information 
the natural assumption would be to say that the sum is comparable with 
$\Delta_{1/m^2}$. Indeed, the origin of $\Delta_{1/m^2}$ can be traced to the
contribution of the highly excited states with the excitation energies of order of $m_c$ while 
the sum under consideration refers to less excitation energies. If some kind of continuity
exists then an assumption
\be
\sum_{m_{D^*}<m_{X_c}<m_{D^*}+\mu} |h_{n}(1)|^2 =(0.5\pm 0.5)\,\Delta_{1/m^2}
\ee
seems reasonable for the range of $\mu \sim 0.5\, m_c$.

Summarizing numerical estimates we get (see Ref.~\cite{bsu} for details) 
\be
h_{A_1}(1)= 0.91 -0.013 \,\,\frac{\mu_\pi^2 -0.5\,{\rm GeV}^2}{0.1\,{\rm GeV}^2}
\pm 0.02_{\rm excit} \pm 0.01_{\rm pert} \pm 0.025_{1/m^3}
\ee
where the uncertainties are marked by the their sources.
Overall,
\be
h_{A_1}(1)= 0.91\pm 0.06\;,
\ee
what leads to the 6\% uncertainty in the $|V_{cb}|$ extraction from the exclusive
$D^*$ production at zero recoil. The theoretical uncertainty  is less by  
about a factor of two in the
determination of  $|V_{cb}|$  from inclusive production~\cite{bsu}. The  difference
is  due to the fact that the inclusive approach is much less sensitive to power
in $1/m_c$ corrections which are the main source of corrections discussed above.
Indeed, for inclusive widths an expansion in powers of $1/m_b$ exists even for a small
mass of final quark.

\section{Subleading Form Factors}
In difference with $h_{A_1}(1)$ form factors $R_1(1)$, $R_2(1)$ and $R_3(1)$ are not
protected against corrections of the first order in 1/m at zero recoil. For this reason they 
can deviate from unity (HQET limit) quite considerably.  Theoretical estimates of
Ref.~\cite{neub94} for $R_1(1)$, $R_2(1)$ are 
\be
 R_1(1)\approx  1 + \frac{m_D* - m_c}{2m_c}+ \frac{m_B* - m_b}{2m_b}\approx 1.35\;;~~~~~~
R_2(1)\approx 0.79\;.
\label{R}
\ee
The theoretical expression for $ R_1(1)$ is presented in somewhat simplified form  to
demonstrate that the  quite considerable deviation from the HQET value comes from the 
$1/m_c$ correction.
The
predictions (\ref{R}) are consistent with CLEO experiment~\cite{CLEO} where form factors were
measured,
\be
R_1=1.18\pm0.30\pm0.12\;; ~~~~~R_2=0.71\pm0.22\pm0.07\;.
\ee

If corrections of the first order in $1/m_c$ are so large,then the natural question to ask is about 
a magnitude of $1/m_c^2$ corrections. The method of sum rules discussed in the previous 
part for $h_{A_1}(1)$ can be applied to get information about other form factors.  Using a set of
sum rules from Ref.~\cite{bsuv} I came to the following estimate for the $1/m^2$ correction
to $R_2(1)$
\be
\left[\Delta R_2(1) \right]_{1/m^2} \approx \frac{\mu_G^2}{m_c^2\left(\frac{m_B}{m_{D^*}}-
1\right)}\left( 1+ \frac{m_c}{3\,m_b}\right) \approx 0.15
\ee
The size of the correction is larger that the one of the first order in $1/m$ given by Eq.~(\ref{R}),
thus the series in powers of $1/m$ does not look as a convergent expansion. The lesson is that one
should be cautious with powers of $1/m_c$.
\section{Some Recent Developments}
This is a brief mention of a  few recent studies relevant to the topic: 
\begin{itemize}
\item{Constraints for form factors following from inclusive sum rules are explored
for different semileptonic decays~\cite{BR}.  It results in upper 
and lower bounds for transitions of $B$ meson to $D$, $D^*$, $\rho$, $\pi$, $\omega$, $K$ and
 } 
\item{Inclusive sum rules following from polarization operator of the semileptonic current
are used to put a new type of bound for the cross channel (like $B\, D$) and in this way to the
form factors~\cite{BGL}.
}
\item{
Extraction of $|V_{ub}|$ from  exclusive modes~\cite{LW}. Using the set of transitions,
$B\ra\rho$, 
$B\ra K^*$,  $D\ra\rho$, $D\ra K^*$ it is possible to diminish considerably a model dependence.
}
\item{
Semileptonic $B$ decays to excited states~\cite{LLSW}. This interesting development is presented
at this conference by Zoltan Ligeti (see his talk).
}
\end{itemize}
\section{Conclusions}
\begin{itemize}
\item{The theory of semileptonic decays is in a good shape}
\item{The main source of theoretical uncertainties is in the $(1/m_c)^n$ corrections, not in the perturbative ones}
\item{Convergence of exclusive and inclusive approaches}
\end{itemize}

\section*{Acknowledgments}
I am thankful to Zoltan Ligeti, Mikhail Shifman, Nikolai Uraltsev and Mikhail Voloshin for
helpful discussions. This work was supported in part by DOE under the grant number
DE-FG02-94ER40823.


\section*{References}
\begin{thebibliography}{99}

\bibitem{HQET}
E. Eichten and B. Hill, {\it Phys. Lett.} {\bf B234} (1990) 511;
\item[] H. Georgi, {\it Phys. Lett.} {\bf B240} (1990) 447.
 
\bibitem{Neub} M. Neubert, 
{\it Phys. Reports} {\bf 245} (1994) 259.
\item[] 
T. Mannel, in: {\it Schladming 1996, Perturbative and nonperturbative aspects of quantum field
theory}  387-428; hep-ph/9606299 

\bibitem{wilson}
K. Wilson, {\it Phys. Rev.} {\bf 179} (1969) 1499;
\item[] K. Wilson and J. Kogut, {\it Phys. Reports} {\bf 12} (1974) 75.

\bibitem{buv}
I. Bigi, N. Uraltsev and A. Vainshtein, {\it Phys. Lett.} {\bf B293} 
(1992) 430; (E) B297 (1993) 477;
\item[] B. Blok and M. Shifman, {\it Nucl. Phys.} {\bf B399} (1993) 441; 459.

\bibitem{bb}
P. Ball and V. Braun, {\it Phys. Rev. } {\bf D49} (1994) 2472.

\bibitem{suv} M. Shifman, N.G. Uraltsev and A. Vainshtein,
{\it Phys.Rev.} {\bf D51} (1995) 2217;  (E) {\bf D52} (1995 ) 3149.

\bibitem{cmu} A. Czarnecki, K. Melnikov and N. Uraltsev, {\it Complete $O(\alpha_s^2)$
Corrections to Zero-Recoil Sum Rules for $B\ra D^*$Transitions}, hep-ph/9706311.

\bibitem{bsu} I. Bigi, M. Shifman and N. Uraltsev {\it Aspects Of Heavy Quark Theory},
hep-ph/9703290

\bibitem{neub94} M. Neubert, {\it Nucl.Phys.} {\bf B416} (1994) 786. 

\bibitem{CLEO} CLEO Collaboration (J.E. Duboscq et al.), {\it Phys.Rev.Lett.} {\bf 76} (1996) 3898.

\bibitem{Luke} M. Luke, {\it Phys. Lett.} {\bf B252} (1990) 447. See earlier applications in:
\item[]  M. Voloshin and M. Shifman, {\it Yad. Fiz.} {\bf 47} (1988) 801 [{\it Sov. J. Nucl. Phys.}
{\bf 47} (1988) 511].

\bibitem{bsuv}I.I. Bigi, M. Shifman, N.G. Uraltsev and A. Vainshtein, {\it Phys. Rev.} {\bf D52}
(1995) 196.

\bibitem{BR} C.G. Boyd and I.Z. Rothstein, {\it Phys.Lett.} {\bf B395} (1997) 96.

\bibitem{BGL} C.G. Boyd, B. Grinstein and R. F. Lebed, {\it  Nucl.Phys.} {\bf B461}
(1996) 493.

\bibitem{LW} Z. Ligeti, I.W. Stewart and M. B. Wise, {\it Comment on $V_{ub}$ from Exclusive
Semileptonic B and D Decays},  hep-ph/9711248. 

\bibitem{LLSW} A.K. Leibovich, Z. Ligeti, I.W. Stewart and M. B. Wise, {\it Semileptonic B decays
to excited charmed mesons}, hep-ph/9705467; {\it Phys.Rev.Lett.} {\bf 78} (1997 ) 3995.

\end{thebibliography}
\end{document}